\documentclass[aps,prl,reprint,groupedaddress]{revtex4-1}

\usepackage[dvips]{graphicx}
\usepackage{amsmath,amsthm,array} 
\usepackage{multirow}
\usepackage{color}
\usepackage{bm}
\usepackage{float}

\usepackage{lipsum} 
\usepackage{ulem} 
\usepackage{multirow}
\usepackage{array}
\newcolumntype{L}[1]{>{\raggedright\let\newline\\\arraybackslash\hspace{0pt}}m{#1}}
\newcolumntype{C}[1]{>{\centering\let\newline\\\arraybackslash\hspace{0pt}}m{#1}}
\newcolumntype{R}[1]{>{\raggedleft\let\newline\\\arraybackslash\hspace{0pt}}m{#1}}
\usepackage{algorithm}
\usepackage{algorithmic}
\usepackage{enumitem}
\setlist[enumerate]{itemsep=0mm}
\usepackage{cleveref}

\usepackage{mdframed}    
\newmdenv[skipabove=6mm, skipbelow=6mm]{kotak}   
\DeclareMathOperator*{\argmax}{arg\,max}

\usepackage{soul}

\begin{document}

\title{Boron cage effects on Nd-Fe-B crystal structure's stability}

\author{Duong-Nguyen Nguyen$^{1,2,*}$, Duc-Anh Dao$^{1}$, Takashi Miyake$^{2,3}$,  Hieu-Chi Dam$^{1,2}$}
\affiliation{
$^{1}$Japan Advanced Institute of Science and Technology, 1-1 Asahidai, Nomi, Ishikawa 923-1292, Japan\\
$^{2}$ESICMM, National Institute for Materials Science 1-2-1 Sengen, Tsukuba, Ibaraki 305-0047, Japan\\
$^{3}$CD-FMat, AIST, 1-1-1 Umezono, Tsukuba, Ibaraki 305-8568, Japan\\
$^{*}$\textit{Corresponding author: nguyennd@jaist.ac.jp}
}

\date{\today}

\begin{abstract}
In this study, we investigate the structure-stability relationship of hypothetical Nd-Fe-B crystal structures using descriptor-relevance analysis and the t-SNE dimensionality reduction method. 149 hypothetical Nd-Fe-B crystal structures are generated from 5967 LA-T-X host structures in Open Quantum Materials Database by using elemental substitution method, with LA denoting lanthanides, T denoting transition metals, and X denoting light elements such as B, C, N and O. By borrowing the skeletal structure of each of the host materials, a hypothetical crystal structure is created by substituting all lanthanide sites with Nd, all transition metal sites with Fe, and all light element sites with B. High-throughput first-principle calculations are applied to evaluate the phase stability of these structures. Twenty of them are found to be potentially formable. As the first investigative result, the descriptor-relevance analysis on  the orbital field matrix (OFM) materials' descriptor reveals the average atomic coordination number as the essential factor in determining the structure stability of these substituted Nd-Fe-B crystal structures. 19 among 20 hypothetical structures that are found potentially formable have an  average coordination number larger than 6.5. By applying the t-SNE dimensionality reduction method, all the local structures represented by the OFM descriptors are integrated into a visible space to study the detailed correlation between their characteristics and the stability of the crystal structure to which they belong. We discover that unstable substituted structures frequently carry Nd and Fe local structures with two prominent points: low average coordination numbers and fully occupied B neighboring atoms. Moreover, there are only three popular forms of B local structures appearing on all potentially formable substituted structures: cage networks, planar networks, and interstitial sites. The discovered relationships are promising to speed up the screening process for the new formable crystal structures.

\end{abstract}



\maketitle

\section{1. Introduction}\label{introduction}

There is a recurring and crucial demand for naturally-sourced and formable materials. The discovery of these stable structures attracts significant attention from various fields. The field of material screening is widely spread from Heusler structures \cite{PhysRevB.95.024411} and rare-earth magnetic structures \cite{Korner16, KORNER18, Jiangang18, Balluff17} to the perovskite structure family \cite{Emery16, MICHALSKY2017124} or topological insulator structures \cite{Yang11, Li_2018}. The major challenge in the discovery of these materials stems from the countless number of hypothetical structures, with only a few likely to be formable. A common hypothesis is that there is a hidden feature or features that cause the instability in the structures. By obtaining these stability-driven features, researchers may inexpensively predict \cite{Oganov19} the stability of newly hypothetical structures and narrow down the area of research thereafter. There are several essential steps for mining these hidden features: (1) generating hypothetical structures along with a reliable stability estimation, (2) modeling the geometrical information of these hypothetical structures using interpretable representation, and (3) building an objective function to analyze the correlations between the geometrical information and the stability.

As a real example, we consider the geometry of permanent magnets. The current strongest permanent magnet is Nd-Fe-B based magnet, whose the main phase is Nd$_2$Fe$_{14}$B. Before the Nd-Fe-B magnet was invented by Sagawa et al. in the early 1980s \cite{Sagawa84, Sagawa86}, the strongest magnet was the Sm$_2$Co$_{17}$ magnet. Due to the need to use iron instead of cobalt, extensive research was conducted in the late 1970s to develop iron-based strong magnets. A potential candidate material was Sm$_2$Fe$_{17}$, or more generally R$_2$Fe$_{17}$ with R denoting the earth element. However, the Curie temperature of R$_2$Fe$_{17}$ is too low to be utilized as a strong magnet at room temperature. The idea of Sagawa et al. was to add a light element to R$_2$Fe$_{17}$ to enhance ferromagnetism by the magneto-volume effect. Boron was chosen as the light element, which leads to the successful development of a strong magnet. Contrary to Sagawa's expectation, however, the main phase of the newly developed strong magnet was not Nd$_2$Fe$_{17}$B$_x$, but the famous Nd$_2$Fe$_{14}$B with a completely different crystal structure. This is a good example of the importance of understanding structure for developing novel materials \cite{Tatetsu18}.

In this research, we investigate  the structure--stability relationship hidden in the screening results of Nd-Fe-B hypothetical structures. Firstly, we collect all possible  LA-T-X crystal structures with lanthanides (LA), transition metals (T), and light elements X = B, C, N, and O from Open Quantum Materials Database \cite{OQMD}. By sharing the same elemental type with Nd-Fe-B structure family, we assume that a number of geometrical skeletons of the LA-T-X crystal structures can be shareable through Nd-Fe-B structures. Therefore, a hypothetical crystal structure of is created by applying the atomic substitution method to the LA-T-X crystal structures.  We replace all lanthanide sites with Nd, all transition metal sites with Fe, and all light element sites with B to create new hypothetical crystal structures (section 2.1). By performing first-principle calculation, we estimate the phase stability (hereinafter referred to as stability) of all these hypothetical structures (section \ref{sec:CH_distance}2.2). In investigating the role of Machine Learning,  new hypothetical structures are first represented using the orbital field matrix (OFM) representation vector. (section 3.1) Relevance analysis is performed to extract the hidden structural descriptors that are essential for determining the stability of the generated Nd-Fe-B crystal structures (section 3.2). Finally, we perform  t-SNE dimensionality reduction method over Nd-based, Fe-based, and B-based local structures to extract correlations between local structure distribution and the structures' stability (section 3.3). 

\section{2. Screening for potentially formable N\MakeLowercase{d}-F\MakeLowercase{e}-B crystal structures}
\subsection{2.1 Creation of new crystal structures}

We collected 5967 crystal structures from the Open Quantum Materials Database (OQMD) \cite{OQMD}(version 1.1) with details calculated formation energies to build the host materials’ dataset denoting by $\mathcal{D}_{\text{LA-T-X}}^{host}$. At the time of conducting this research, the Open Quantum Materials Database contains more than 637,000 materials with DFT calculated thermodynamic and structural properties. In this database, roughly 30\% of structures in the OQMD are stable, and this portion is reducing yearly by the appearance of more calculated results of hypothetical structures  \cite{OQMD}.  All the structures in $\mathcal{D}_{\text{LA-T-X}}^{host}$ comprise lanthanide (LA), transition metal (T), and light (X) elements. The LA elements are $\{$Y, La, Ce, Pr, Nd, Pm, Sm, Eu, Gd, Tb, Dy, Ho, Er, Tm, Yb, and Lu$\}$. The T elements are $\{$Ti, V, Cr, Mn, Fe, Co, Ni, Cu, Zn, Y, Zr, Nb, Mo, Tc, Ru, Rh, Pd, Ag, Cd, Hf, Ta, W, Re, Os, Ir, Pt, Au, and Hg$\}$. The X elements are $\{$H, B, C, N, and O$\}$.   Any crystal structure in $\mathcal{D}_{\text{LA-T-X}}^{host}$ included one or two rare-earth metals, one or two transition-metals, and one light element. From $\mathcal{D}_{\text{LA-T-X}}^{host}$,  we denoted a subset of all crystal structures comprising Nd, Fe, and B as $\mathcal{D}_{\text{Nd-Fe-B}}^{host}$.

In this research, elemental substitution method was applied on the crystal structures in $\mathcal{D}_{\text{LA-T-X}}^{host}$ to generate new hypothetical structures consisting of Nd, Fe, and B. We substituted all lanthanide sites of a given host crystal structure with Nd, all transition metal sites with Fe, and all light element sites of  with B to generate a new hypothetical Nd-Fe-B structures. The newly substituted structures were compared to each other and to the crystals in the $\mathcal{D}_{\text{LA-T-X}}^{host}$ dataset to avoid  duplication. In the first step, we remove all possible structural duplications using the comparison procedure proposed by qmpy (python application programming interface of OQMD) \cite{OQMD}.  In the first step of the qmpy procedure, lattice parameters of all  hypothetical structure pairwises in the reduced primitive cells are compared. Secondly, the structural comparison proposed by qmpy compared the internal coordinates of the structures by examining all the possible structure rotations and translations to map atoms of the same species into one another within a given deviation. In this step, any two structures that have the same deviation amount of lattice parameters and angles smaller than 0.1 percent were consided identical. In the last step, we used our developed OFM descriptor (section \ref{sec:OFM}3.1) for removing duplications. Two structures were set identical if the $L_{2}$ norm of the difference in the OFM descriptor was less than $10^{-3}$. It is worth noticing that two structures that have the same shape yet are slightly different in size are considered identical. Finally, we obtained a dataset $\mathcal{D}_{\text{Nd-Fe-B}}^{subst}$ containing 149 new nonoptimized Nd-Fe-B crystal structures.

\subsection{2.2 Assessment of phase stability}\label{sec:CH_distance}
First-principle calculations based on density functional theory (DFT) \cite{KS-DFT, HK-DFT} are broadly accepted in computational materials science. DFT calculations accurately determine the formation energy of materials, which is used to build phase diagrams for systems of interest. Consequently, the phase stability of a material ---in other words, the decomposition energy of a given material or CH-distance is obtained through the convex-hull analysis of phase diagrams and the decomposition of the material into other phases. We used the formation energy obtained from OQMD \cite{OQMD, OQMD1} of $\mathcal{D}_{\text{LA-T-X}}^{host}$ to build phase diagrams and calculate the CH-distance. The CH-distance of a material is determined as follows:

\begin{equation}
	\label{eq:convex_hull}
	\Delta E = \Delta E_f - E_H,
\end{equation}
where $\Delta E_f$ is the formation energy, and $E_H$ is determined by projecting from the chemical composition position to an endpoint appearing on the convex hull facets. Details of the algorithm for determining these convex hull facets from hull points and $E_H$ are given in \cite{Barber96, OQMD}. Hence, we consider the CH distance $\Delta E$ to measure the phase stability of a material. A material that lies below or on the CH surface, $\Delta E = 0$, is potentially formable. In contrast, a material associated with $\Delta E > 0$ is unstable. A material associated with $\Delta E$ lightly above the CH surface is considered a metastable phase. Referring to the prediction accuracy of formation energy ($\sim$ 0.1 eV/atom by OQMD \cite{OQMD}), we defined all the materials with $\Delta E \leq 0.1$ eV/atom as potentially formable materials and as unstable materials otherwise. Following this definition, $\mathcal{D}_{\text{LA-T-X}}^{host}$ could be divided into subsets $\mathcal{D}_{\text{LA-T-X}}^{host\_stb}$ and $\mathcal{D}_{\text{LA-T-X}}^{host\_unstb}$ for potentially formable crystal structures and unstable crystal structures, respectively.

$\mathcal{D}_{\text{Nd-Fe-B}}^{host}$ holds 35 Nd-Fe-B crystal structures that are used as references to build the Nd-Fe-B phase diagram. Among them, there are seven ternary crystal structures. To verify the reliability of the dataset in constructing the phase diagram and the definition of structure stability, we removed each ternary crystal then used the remaining crystals in $\mathcal{D}_{\text{Nd-Fe-B}}^{host}$ to determine its corresponding stability value. As a result, among the seven ternary materials, NdFe$_4$B$_4$ and Nd$_5$Fe$_2$B$_6$ are two formable ternary crystals with $\Delta E = 0.0$. Moreover, one material, NdFe$_{12}$B$_6$, is potentially formable (metastable) with its stability less than $0.1$ eV/atom. Table I in the Supplementary Materials shows details about the stability of all these crystal. It should be noted that the important magnetic material, Nd$_2$Fe$_{14}$B, did not exist in the OQMD database when we conducted this study. Using the DFT calculated formation energy of -0.057 (eV/atom), the $\Delta E^{DFT}$ of is Nd$_2$Fe$_{14}$B $1.4 \times 10^{-4}$ eV/atom. This result indicates that Nd$_2$Fe$_{14}$B is in the stable phase.  To conclude, we evaluated the stability definition by determining the stability of these experimentally manufactured structures. 

We followed the computational settings of OQMD \cite{OQMD, OQMD1} to determine the formation energy of the hypothetical Nd-Fe-B crystal structures in $\mathcal{D}_{\text{Nd-Fe-B}}^{subst}$. The calculations were performed using the Vienna ab initio simulation package (VASP) \cite{vasp1,vasp2,vasp3,vasp4} by utilizing projector-augmented wave (PAW) method potentials \cite{paw1, paw2} and the Perdew--Burke--Ernzerhof (PBE)\cite{pbe} exchange-correlation functional.

We employed DFT+U for  Fe, and all calculations were spin-polarized with the ferromagnetic alignment of the spins. Initial  values of the magnetic moment were set at 5, 0, and 0 $\mu_B$ for Fe, Nd, and B respectively.  For any new  structure, we followed three optimization steps: 'coarse relax',  'fine relax', and  'standard' procedures of the OQMD for  coarse optimization, fine optimization, and a single point calculation, respectively. The number of  k-points per reciprocal atom (KPRA) was set as 4000, 6000, and 8000 for the 'coarse relax', 'fine relax', and  'standard' optimization steps respectively. We used a cutoff-energy value at 520 eV for all calculations. The total energy obtained from the "standard" calculation step was used to estimate the formation energy, $\Delta E^{DFT}_f$. In the last step, the CH distance of a new structure is estimated from equation \ref{eq:convex_hull}.  
 
\begin{figure}[t]
	\includegraphics[width=0.5\textwidth]{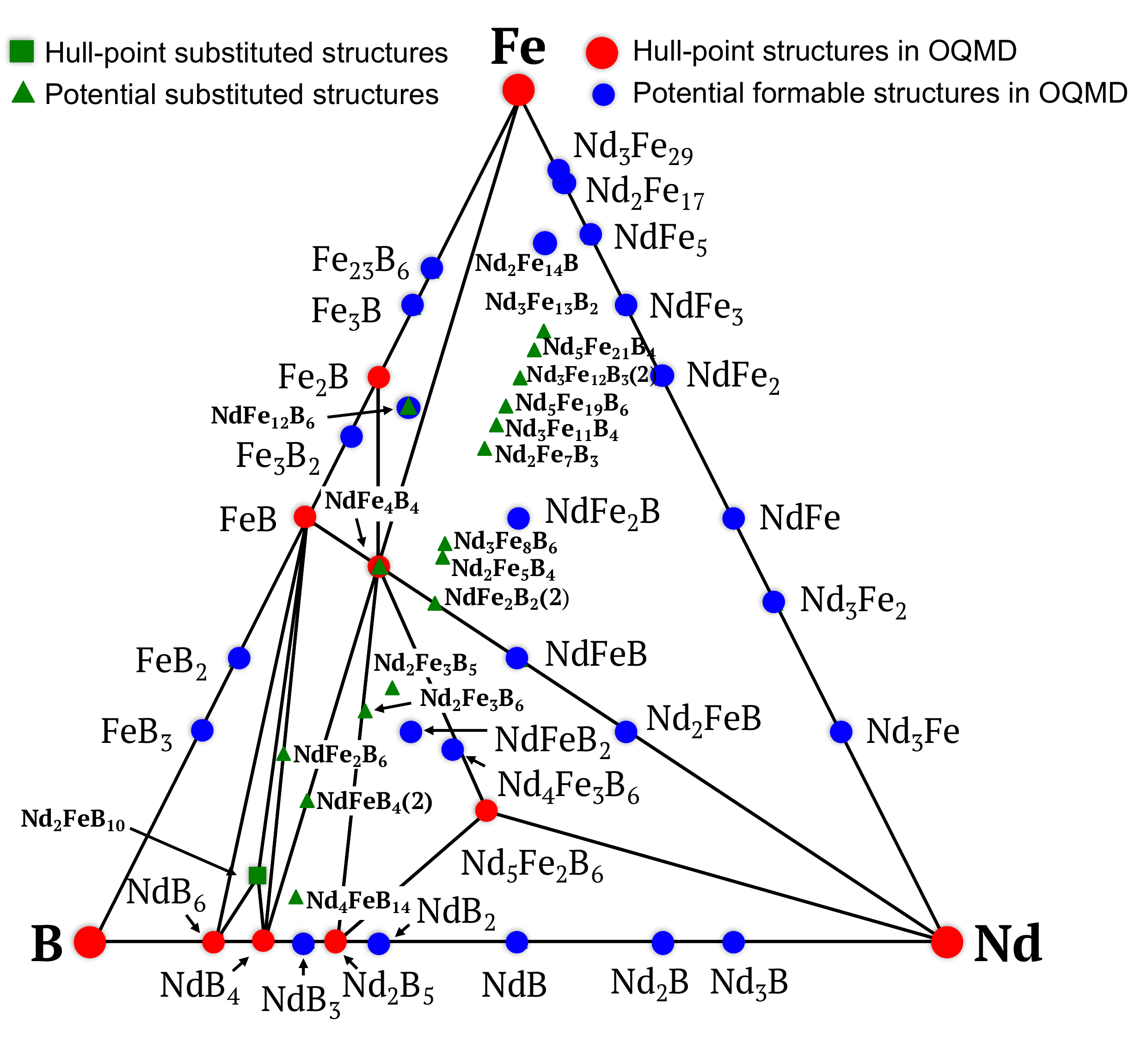}
    \caption{\label{fig:phase_diag} Phase diagram of Nd-Fe-B including materials obtained from OQMD (circles) and 20 new substituted structures that confirm it is potentially formable (green points). OQMD hull points are denoted by red circles, and possible hull point from substituted structures are denoted by green square. The total number of disparate structures with the same chemical composition is shown in parentheses.}
\end{figure}

Finally, we obtained 20 new Nd-Fe-B crystal structures that are not in $\mathcal{D}_{\text{LA-T-X}}^{host}$, in which the CH distance of the corresponding optimized crystal structure was less than 0.1 eV. Figure 2 in the Supplementary Materials shows details of their geometrical shapes. These crystals originate from different host crystal structures with different skeletons. Note that we found one structure, Nd$_2$FeB$_{10}$, with stability less than $-0.01$ eV/atom. Accordingly, this structure is also used as a reference to construct the Nd-Fe-B phase diagram. Among the 20 new Nd-Fe-B structures, there are three pairs of discriminated structures sharing the same chemical compositions (NdFe$_2$B$_{2}$, NdFeB$_{4}$ and NdFe$_4$B).  Table II in the Supplementary Materials shows details about the DFT calculated formation energy and the stability of all these crystals. Figure \ref{fig:phase_diag} shows the phase diagram of the Nd-Fe-B crystals, including the 20 new substituted crystal structures. 

\begin{figure*}[t]
	\includegraphics[width=1.0\textwidth]{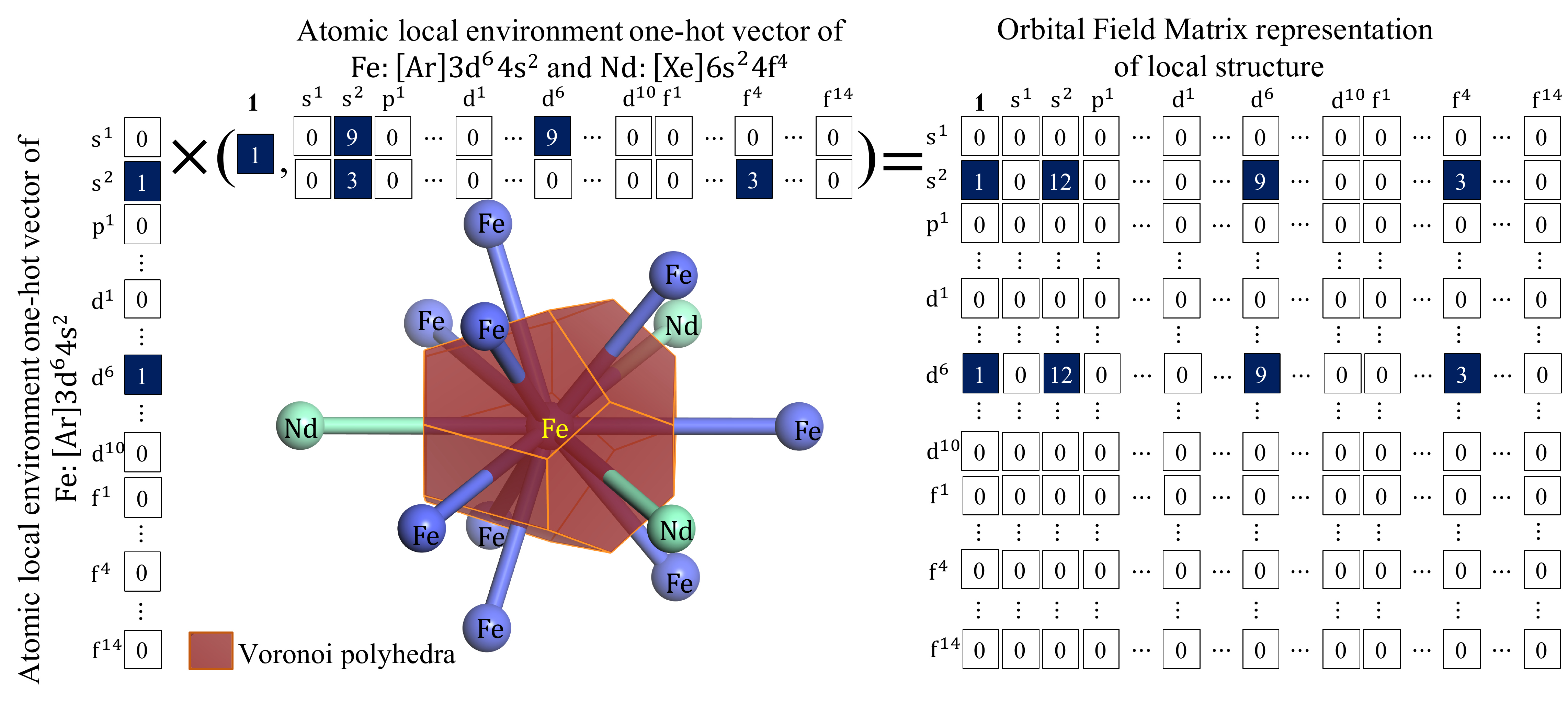}
    \caption{\label{fig:ofm} Orbital Field Matrix (OFM) representation of a local structure with the center Fe atom surrounding by nine Fe other neighbors and three Nd neighbors. The solid angle creating by Voronoi polyhedra faces are counted as weights of corresponding neigbors to the center.}
\end{figure*}
\section{3. Mining structure--stability relationship of N\MakeLowercase{d}-F\MakeLowercase{e}-B crystal structures}\label{sec:ML_models}
\subsection{3.1 Materials representation}\label{sec:OFM}

In this study, we employed the OFM  \cite{ofm, ofm1} to convert the information of materials to the description vector. The OFM descriptors were constructed using the weighted product of the one-hot vector representations, $\vec{O}$, of atoms. Each vector $\vec{O}$ is filled with zeros, except those representing the electronic configuration of the valence electrons of the corresponding atom. The OFM of a local structure, named $\Theta$, is defined as follows:

\begin{equation}
    \label{eq:ofm_no_d}
     \Theta = \vec{O}^{\top}_{central} \times \left(1.0, \sum_k\frac{\theta_k}{\theta_{max}}\vec{O}_{k}\right),
\end{equation}

where $\theta_k$ is the solid angle determined using the face of the Voronoi polyhedra between the central atom and the index $k$ neighboring atom; $\theta_{max}$ is the maximum solid angle between the central atom and neighboring atoms. By removing the distance dependence in the original OFM formulation \cite{ofm, ofm1}, we focused exclusively on the coordination of valence orbitals and the shape of the crystal structures. The mean over the local structure descriptors is used as the descriptor of the entire structure:
\begin{equation}
  \label{eq:6} 
\quad OFM_{p} = \frac{1}{N_{p}} \sum_{l=1}^{N_{p}} \Theta_{p}^{l},
\end{equation}
where $p$ is the structure index, and $l$ and $N_p$ are the local structure indices and the number of atoms in the unit cell of the structure $p$, respectively. A complete representation of the OFM matrix has 1056 dimensions. However, under a given investigation system, a large number of zero terms appearing in all structures’ representation due to the absence of other elements. In this investigation, Nd-Fe-B structures with three atomic types require 19 non-zero OFM variables. Details illustration of OFM descriptor are shown in Figure \ref{fig:ofm}

 Note that, owing to the designed cross product between the atomic representation vectors of each atom, each element in the matrix represents the average number of atomic coordinates for a certain type of atom. For example, an element of a descriptor obtained by considering the product of the $d^6$ element of the center atom representation and the $f^4$ element of the environment atom representation, denoted as  {$\left(d^6, f^4\right)$}, shows an average coordination number of $f^4$ (Nd) sites surrounding all $d^6$ (Fe) sites. As the term $s^{2}$ appears at all descriptors for Fe, Nd, and B sites, the element {$\left(s^2, s^2\right)$} represent the average coordination number of a given structure. All these OFM elements make it possible to investigate the structure--stability relationship.

\subsection{3.2 Descriptor relevant analysis}\label{sec.method_relevant}

In this section, we focused on $\mathcal{D}_{\text{Nd-Fe-B}}^{host}$ and evaluated the relevance \cite{Nguyen_2019, feature_selection_Yu:2004, feature_selection} of each element in the OFM descriptor regarding the formation energy of the crystal structure. We utilized the change in prediction accuracy when removing or adding an element (from the full set of elements in the OFM) to search for the elements that were strongly relevant \cite{Nguyen_2019, Dam18} to the formation energy (i.e., CH distance and phase stability) of the Nd-Fe-B crystal structures.

In detail, for a given set $S$ of descriptor elements, we defined the prediction capacity $PC(S)$ of $S$ by the maximum prediction accuracy that a prediction model could achieve by using the variables in a subset $s$ of $S$ as follows:

\begin{equation}
PC(S) = \max_{\forall s \subset S} R^2_s; \\  s_{PC} = \argmax_{\forall s \subset S} R^2_s, 
\label{eq.PAS}
\end{equation}
where $R^2_s$ is the value of the coefficient of determination $R^2$ \cite{Kvalseth85} achieved by the prediction model using a set $s$ as the independent variables. $s_{PC}$ is the subset of $S$ that yields the prediction model having the maximum prediction accuracy. In this study, we used kernel ridge regression (KRR)\cite{ML} as the prediction model. 

$S_i$ is denoted as set of descriptor elements after removing an element $x_i$ from the full descriptor element set $S$; $S_i = S - \{x_i$\}.  A descriptor element is strongly relevant if and only if:

\begin{equation}
PC(S)-PC(S_i) = \max_{\forall s \subset S} R^2_s - \max_{\forall s \subset S_i} R^2_s > 0.
\label{eq.strong}
\end{equation}

\begin{figure}[t]
\centering
  \includegraphics[width=0.95\linewidth]{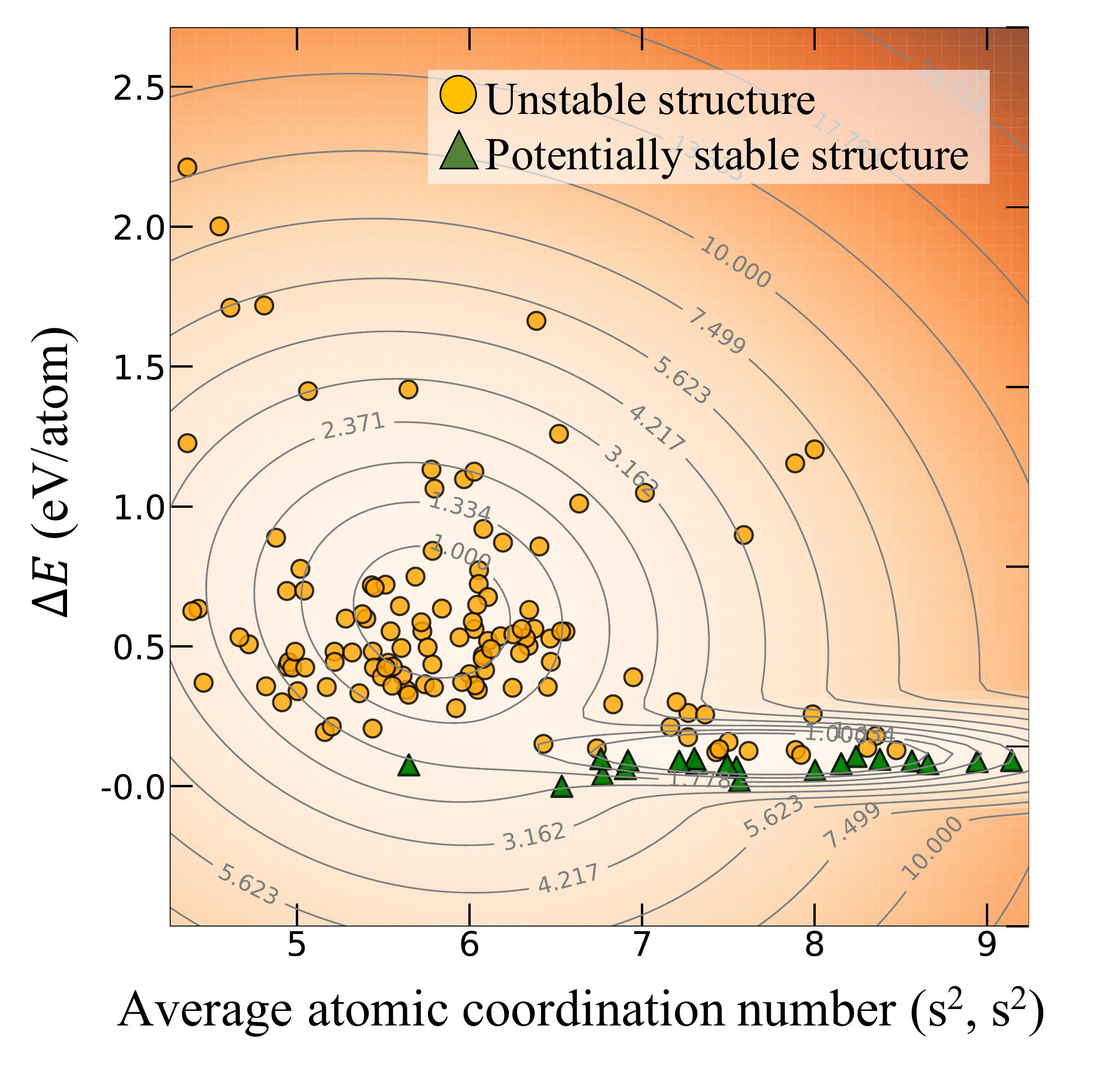}
\caption{Dependence of structure's stability on the OFM variables $(s^2, s^2)$\label{fig:Fig6} of Nd-Fe-B substituted structures. The $(s^2, s^2)$ feature represents for the average atomic coordination numbers. Potentially formable (unstable) structures are denoted by a green triangle (orange circle). Contour lines show surfaces of  iso-density distribution.}
\end{figure} 

By applying the descriptor relevant analysis in predicting the formation energy of the crystal structures in $\mathcal{D}_{\text{Nd-Fe-B}}^{host}$ it can be claimed that the descriptor $(s^2, s^2)$ is a strong relevant descriptor. Details of the result are shown in Supplementary materials. 

In Figure \ref{fig:Fig6}, we show the dependence of the CH-distance, $\Delta E$ on the average atomic coordination number $(s^2, s^2)$ of crystal structures in $\mathcal{D}_{\text{Nd-Fe-B}}^{subst}$ with Gaussian kernel density estimation implemented in sklearn.mixture.GaussianMixture\cite{scikit-learn}. The distribution of $\mathcal{D}_{\text{Nd-Fe-B}}^{subst}$ shown in this space is a mixture of two main  components. The larger distribution component is located in the region with $(s^2, s^2) < 6.5$, whereas the other is located in the region with $(s^2, s^2) \geq 6.5$. Contour lines show surfaces of iso-density distribution. We inferred the existence of two distinct groups of substituted crystal structures. The first group contained structures with average atomic coordination numbers lower than 6.5. The second group contained structures with average atomic coordination numbers higher than 6.5. Interestingly, almost all the new, potentially formable crystal structures -- 19 among 20 substituted structures belonged to the second group. As mentioned, it is evident that the stability mechanism of Nd-Fe-B substituted structures is closely related to the average atomic coordination number.

\begin{figure}[t]
	\includegraphics[scale=0.4]{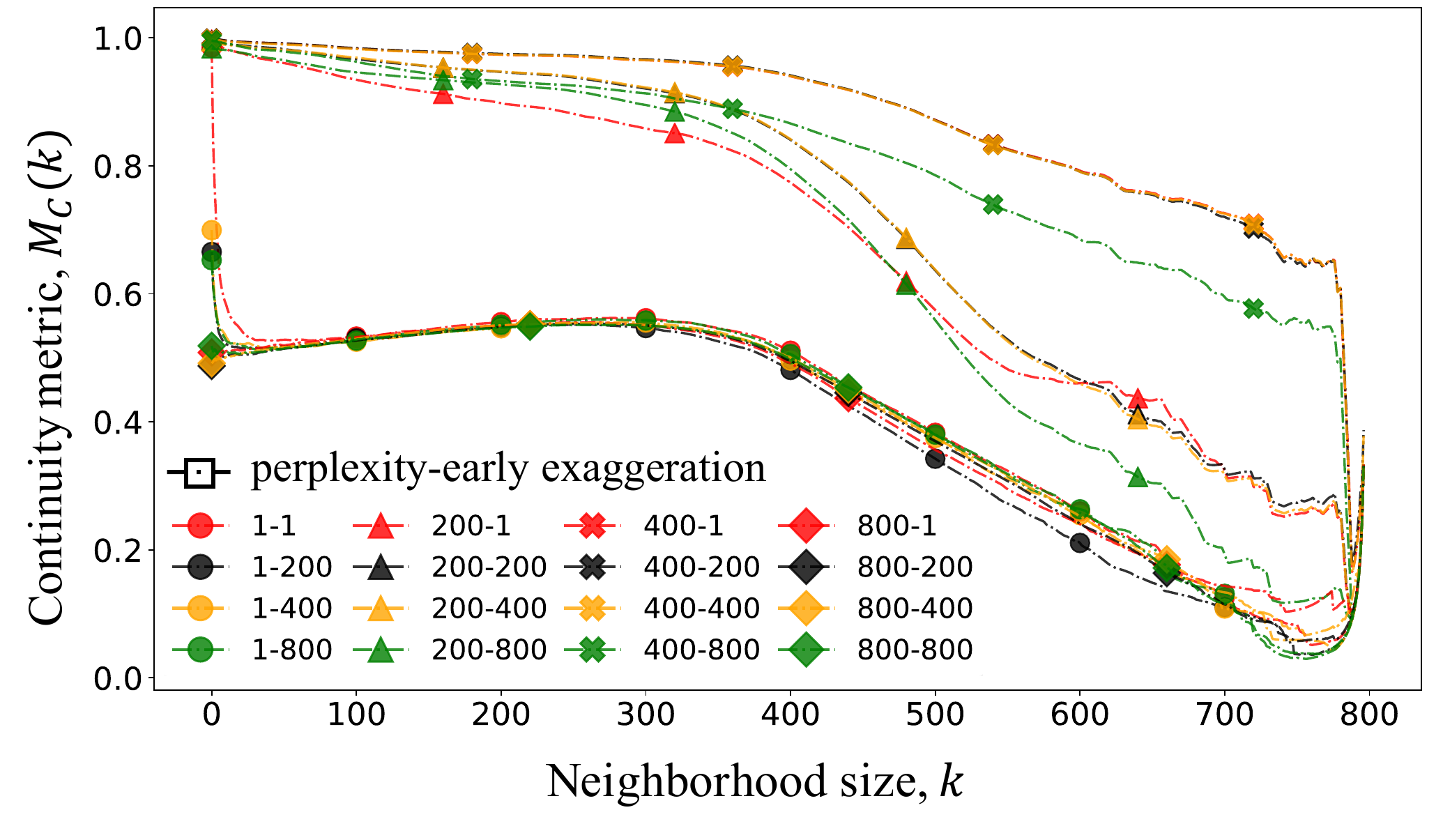}
    \caption{Continuity metric $M_{C}(k)$ of the t-SNE method in embedded Nd local structure data set to 2D  space with different perplexity and early exaggeration parameters.} \label{fig:Nd_continuity}
\end{figure}

\begin{figure*}
	\includegraphics[scale=0.45]{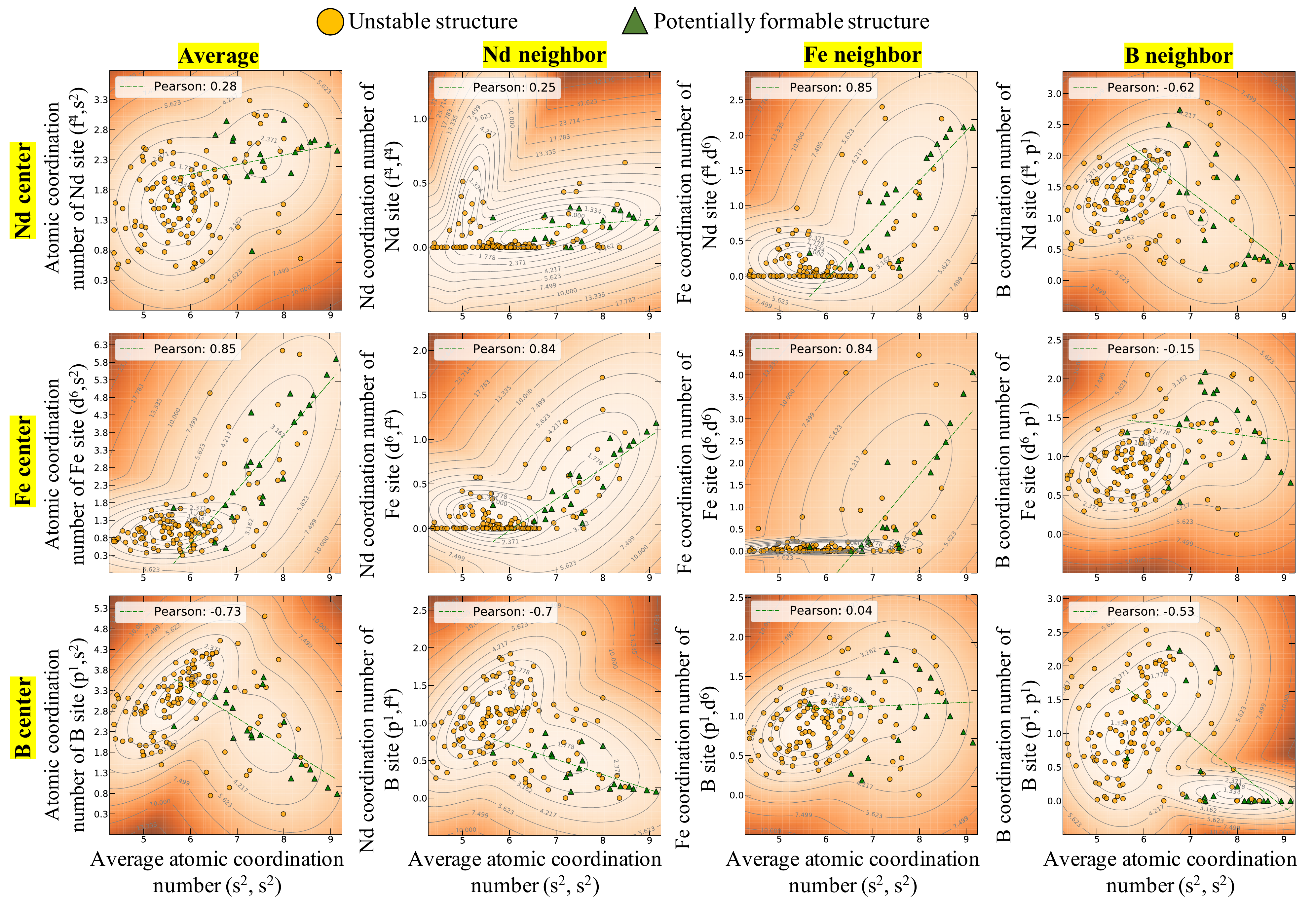}
    \caption{\label{fig:Distribution}Density distribution of newly generated Nd-Fe-B crystal structures in two-dimensional space obtained using selected OFM elements. The blue and red solid circles represent the unstable and potentially formable crystal structures verified by DFT calculations, respectively. Contour lines show the isodense  surface of the distribution.}
\end{figure*}

Next, we investigated the distributions of all the substituted structures in the space of  $(s^2, s^2)$ and the remaining set of OFM descriptors. Based on the fact that each OFM descriptor shows a particular description of coordination number,  we categorized the remaining OFM descriptors according to center-based and neighbor-based components. For instance, the OFM descriptor $(d^6, f^4)$ represents the coordination number of Nd (neighbor) surrounding Fe (center) sites.  All distributions of substituted structures regarding the $(s^2, s^2)$ descriptor and the set of remaining descriptors are summarized in Fig.  \ref{fig:Distribution}. In figure \ref{fig:Distribution}, one might notice that all the distributions were relatively similar to the distribution over $\Delta E$ and $(s^2, s^2)$ spaces, figure \ref{fig:Fig6}. There were two distinguishable groups with one located in the region of $(s^2, s^2) < 6.5$ and the other located in the region of $(s^2, s^2) \geq 6.5$. As previously mentioned, almost all the potentially formable structures appear in the second group, with the average coordination number $(s^2, s^2) \geq 6.5$. However, in the group of these potentially formable structures, different elemental coordination numbers exhibit different correlations with $(s^2, s^2)$. We used Pearson score \cite{Pearson} to measure these linearity correlations qualitatively. The Pearson score has a value between [-1, +1] with -1 and +1 represent for total negative and positive linear correlation, respectively. In practical use, one might consider the absolute value of the Pearson score larger than 0.8 as an existence of linear correlation, otherwise as non-existence of linear correlation. The investigation results are summarized in table \ref{tab:Distribution}.

\begin{table}[t]
\caption{\label{tab:Distribution} Pearson correlation between $(s^2, s^2)$ and other elemental OFM descriptors in the group of potentially formable structures.}
\begin{tabular}{C{18mm}C{15mm}C{15mm}C{15mm}C{15mm} } 
\hline\hline
 	                     & Average & Nd neighbor & Fe neighbor &  B neighbor   \\   \hline \hline                            
Nd center 	& 0.28  & 0.25  & 0.85  & -0.62   \\ \hline
Fe center 	& 0.85 & 0.84 & 0.84 & -0.15   \\ \hline
B center  	& -0.73 & -0.70 & 0.04 & -0.53   \\ \hline

\hline
\end{tabular}
\end{table}

In the group of substituted structures which are potentially formable, $\mathcal{D}_{\text{Nd-Fe-B}}^{subst-stable}$, by increasing $(s^2, s^2)$, descriptors $(d^6, f^4)$ and  $(d^6, d^6)$ show upward trends with Pearson coefficients of 0.84. The descriptor $(d^6, f^4)$ ($(d^6, d^6)$)  represents the average of Nd (Fe) coordination numbers surrounding Fe-center sites. In addition, the average Fe coordination number surrounding Nd-center sites shows a positive correlation with $(s^2, s^2)$ for a Pearson coefficient of 0.85. In contrast, by increasing the $(s^2, s^2)$ value,  Nd and B coordination numbers of B-center sites show apparent downward trends with Pearson coefficients of -0.7 and -0.53, respectively. To conclude, for potentially formable structures in $\mathcal{D}_{\text{Nd-Fe-B}}^{subst-stable}$, by increasing the average coordination number $(s^2, s^2)$, the coordination numbers of Nd and Fe elements surrounding Fe sites and Fe elements surrounding Nd sites also increase. In contrast,  Nd-Fe-B potentially formable structures with  a larger average coordination number possess their B sites with a lower coordination number. All these correlations are summarized in table \ref{tab:Distribution} and figure \ref{fig:Fig6} .

\subsection{3.3 Local structure distribution}\label{sec.local_str_analysis}
In the previous section, we investigated the structure-stability relationship of substituted structures that are potentially formable. Correlations of different coordination numbers regarding the strong relevant feature, $(s^2, s^2)$, are discussed.  In this section, we investigate distributions of  Nd, Fe, and B local structures emphasizing on the stabilty of corresponding substituted structures on t-SNE embedding spaces \cite{tsne08}.

The OFM description vector we used in this work (section 3.1) has 1056 dimensions. By omitting constant variables, in $\mathcal{D}_{\text{Nd-Fe-B}}^{subst-stable}$, there are 19 remaining variables that need to represent these local structures.  To visualize the distributions of local structures, a dimensionality reduction method is necessary. Herein, we employed t-distributed stochastic neighbor embedding (t-SNE) \cite{tsne08}. This technique maps a set of high-dimensional data points to a lower-dimensional space (typically two), such that ideally, close neighbors remain close, and separate data points prevail distantly. Roughly speaking, the algorithm initially samples all local structures on a 2D space, at random positions and then moves these local structures gradually, aiming to minimize derivations between all pairwise distances in the 2D plane and pairwise distances in the original space. The most critical parameter of t-SNE, called perplexity, controls the width of the Gaussian kernel  \cite{tsne08}.  It also effectively governs similarities between local structures and how many of its nearest neighbors each point is attracted to. In other word, the optimal parameter perplexity balance attention between local and global aspects of our data \cite{wattenberg2016how}. The second important parameter of t-SNE is the early exaggeration. This parameter controls the work of forming cluster at initial steps by forcing early clusters tightly knit together. Therefore, there is more empty space to help clusters moving around relative to others to archive an optimal global organization. Details explanation and practical effect in using these parameters are shown in the original paper \cite{tsne08} and \cite{wattenberg2016how, Kobak19}. To obtain distributions in the best match with the original distribution, we evaluate the embedding results taken from the t-SNE method with a designed score and then select the optimal one.

To quantify the quality of a given embedding, we use the two following metrics: trustworthiness ($M_{T}$) and continuity ($M_{C}$) \cite{Venna06, Lee09}. These two metrics were designed to evaluate the information loss of all dimensionality reduction method. In detail, the $M_{T}$ trustworthiness score and $M_{C}$ continuity score are defined as follows:

\begin{eqnarray}
M_T(k) = 1 - {{2}\over{Z}} \Sigma_{i=1}^{N} \Sigma_{j \in U_k(i)} (r(i, j) - k) \\
M_C(k) = 1 - {{2}\over{Z}} \Sigma_{i=1}^{N} \Sigma_{j \in V_k(i)} (\hat{r}(i, j) - k)
\end{eqnarray}

where $Z = Nk(2N-3k -1)$ with $k$ is the size of the neighborhood; $N$ is the total number of local structures. $r(i, j)$ and $\hat{r}(i, j)$  are the rank of local structure $j$ in the ordering according to the distance from $i$ in the original and embedding space, respectively. $U_k(i)$ is the set of local structures that are approximately size $k$ of the local structure $i$ in the embedding space but not in the original space. $V_k(i)$ is the set of local structures that are approximately size $k$ of the local structure $i$ in the original space but not in the embedding space. In comparison, $M_{T}(k)$ measures the degree of reliability exhibited when local structures with further entrances join together in the embedding space, whereas  $M_{C}(k)$ measures the degree of continuity exhibited when local structures with original  locality are pushed farther away in the embedding space. $M_{C}$ and $M_{T}$ metrics are all in range from 0.0 to 1.0. The optimal value 1.0 indicates the most reliable for embedding methods as the distance metric on the embedding space totally matches the distance metric on the original space. On the other hand, the minimum value 0.0 indicates the least reliable results. Based on this interpretation, we performed grid-search parameters and then raised the optimal t-SNE's parameters by maximizing the simple average of $M_{T}$ and $M_{C}$. In the grid-search,  perplexity and early exaggeration parameters vary from 1 to 1000.

\begin{figure*}[t]
	\includegraphics[scale=0.7]{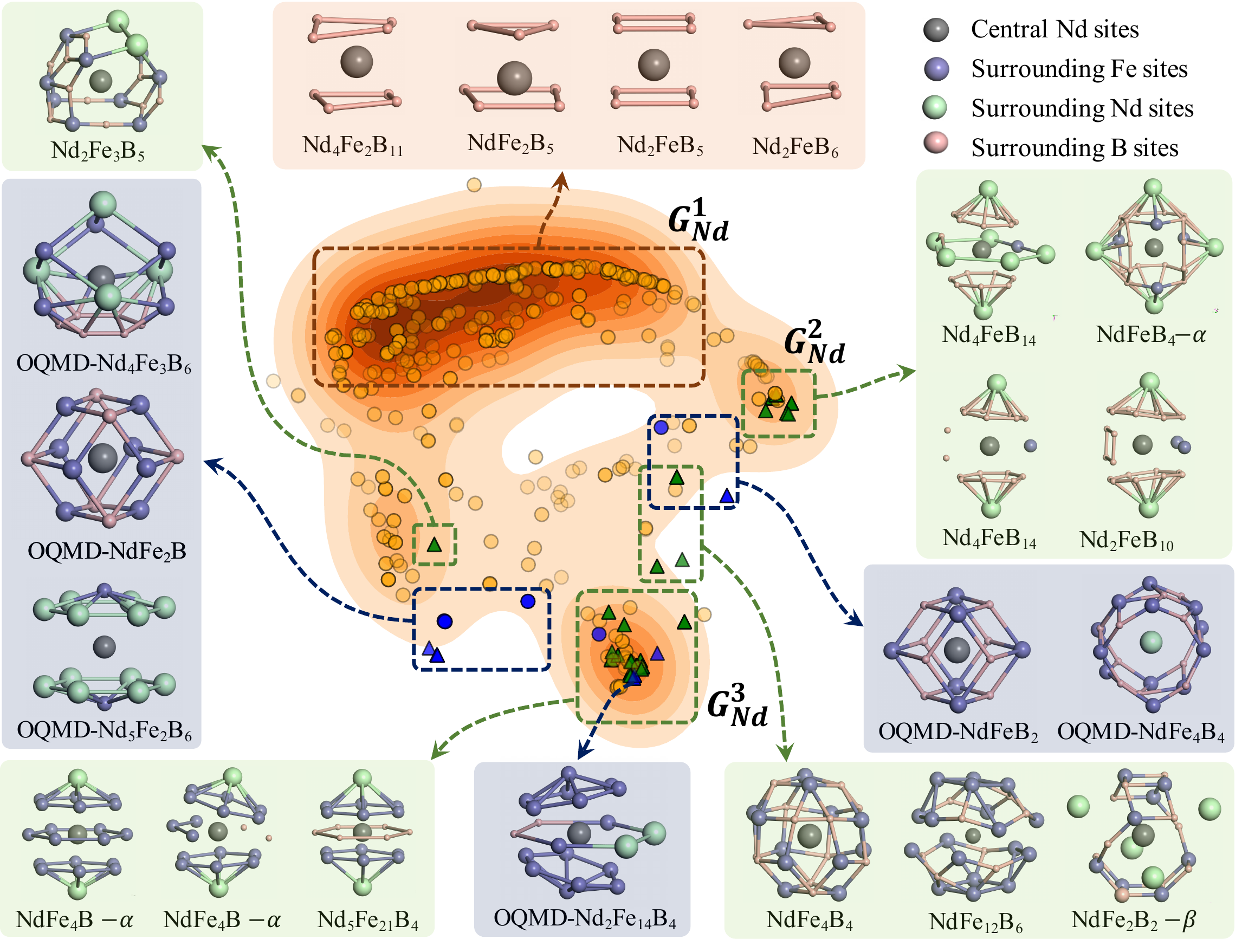}
    \caption{Distribution of Neodymium local structures collecting from all the Nd-Fe-B substituted crystals and Nd-Fe-B crystals from OQMD over t-SNE embedding space. Neodymium local structures gathered from substituted crystal structures which are potentially formable (unstable)  are denoted by green triangles (orange circles). Neodymium local structures gathered from OQMD crystal structures which are potentially formable (unstable)  are denoted by blue triangles (blue circles). Contour lines show surfaces of  iso-density distribution.} \label{fig:local_Nd}
\end{figure*}

Figure \ref{fig:Nd_continuity} shows results of the continuity  $M_{C}$ metric in searching for the optimal parameters of Nd local structure data. The $M_{C}$ metric generally starts at the maximum value of 1.0 with $k=0$ neighbor and then gradually decreases when evaluating larger $k$ neighbors. As shown in this figure, the embedding results using a group of parameters with a perplexity of 400 exhibit a maximum of $M_{C}$ values for all $k$. Finally, we obtain the optimal perplexity (early exaggeration) parameter for Nd, Fe, and B local structures distribution as follows: 400 (400),  800 (200), and 200 (100), respectively.

\begin{figure*}[t]
	\includegraphics[scale=0.8]{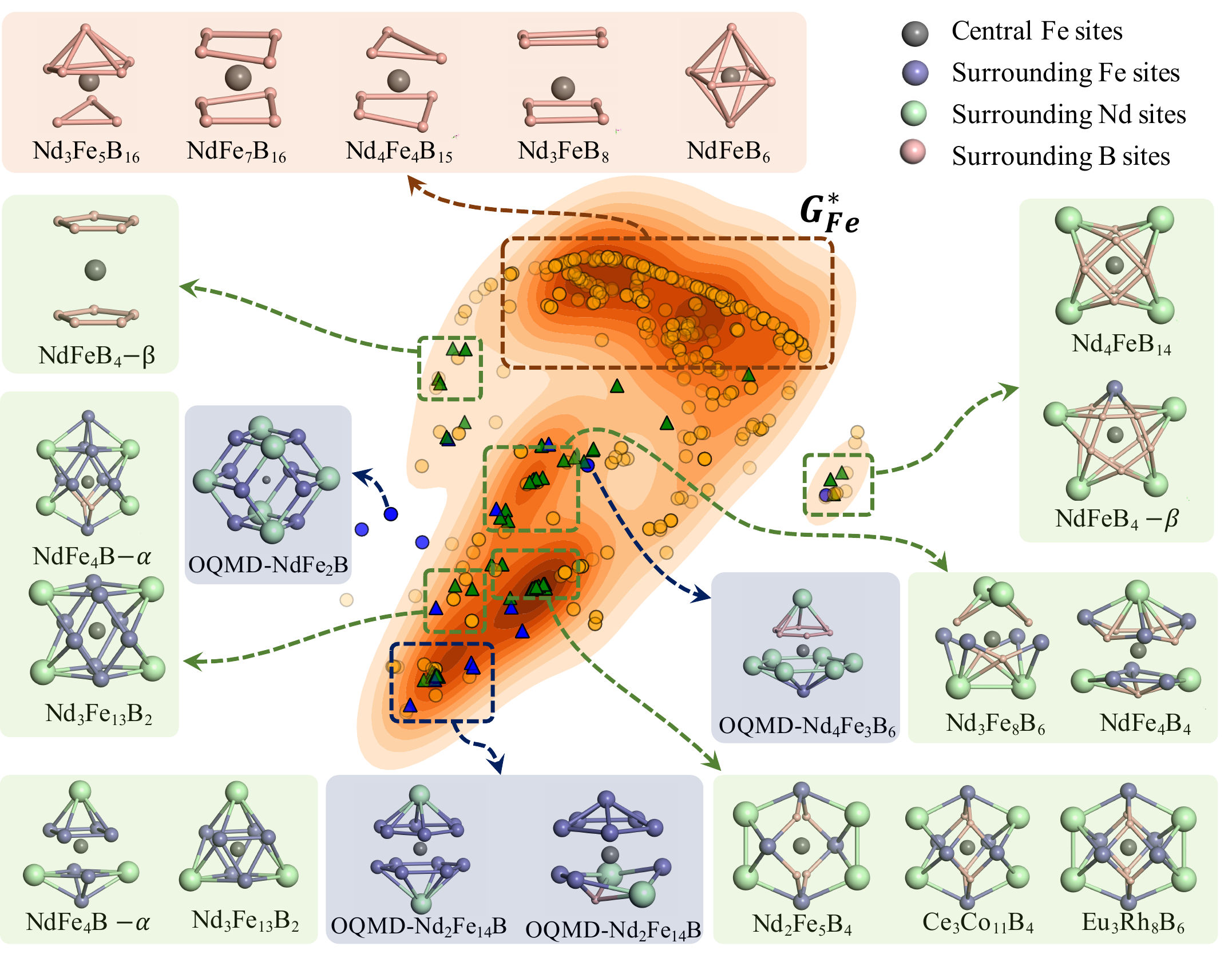}
    \caption{Distribution of Iron local structures collecting from all the Nd-Fe-B substituted crystals and Nd-Fe-B crystals from OQMD over t-SNE embedding space. Iron local structures gathered from substituted crystal structures which are potentially formable (unstable)  are denoted by green triangles (orange circles). Iron local structures gathered from OQMD crystal structures which are potentially formable (unstable)  are denoted by blue triangles (blue circles). Contour lines show surfaces of  iso-density distribution. } \label{fig:local_Fe}
\end{figure*}

\begin{figure*}[t]
	\includegraphics[scale=0.7]{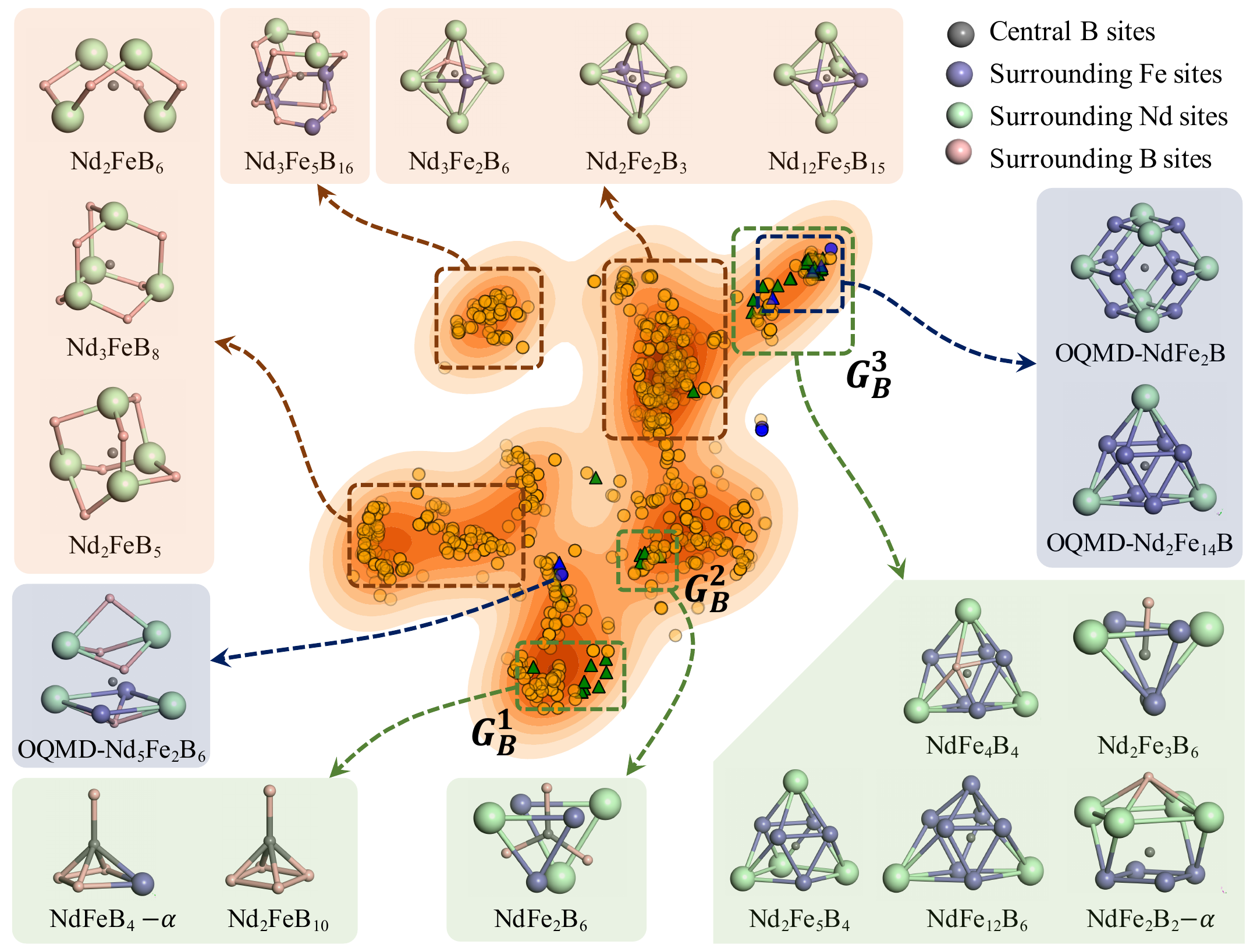}
    \caption{Distribution of Boron local structures collecting from all the Nd-Fe-B substituted crystals and Nd-Fe-B crystals from OQMD over t-SNE embedding space. Boron local structures gathered from substituted crystal structures which are potentially formable (unstable)  are denoted by green triangles (orange circles). Boron local structures gathered from OQMD crystal structures which are potentially formable (unstable)  are denoted by blue triangles (blue circles). Contour lines show surfaces of  iso-density distribution. } \label{fig:local_B}
\end{figure*}

Figures \ref{fig:local_Nd}, \ref{fig:local_Fe}, and \ref{fig:local_B} show the distributions of Nd, Fe, and B local structures respectively by applying the t-SNE dimensionality reduction method with optimal parameters. Local structures gathered from potentially formable substituted structures are denoted by green triangles. In contrast, orange circles are used to denote  local structures gathered from unstable substituted structures. We also project the local structures of all ternary materials obtained from OQMD as references. These ternary materials were estimated the structure stability shown in Sect. 2.2 and also used to construct the Nd-Fe-B phase diagram shown in Fig. \ref{fig:phase_diag}. In figures \ref{fig:local_Nd}, \ref{fig:local_Fe}, and \ref{fig:local_B}, local structures gathered from stable material structures in OQMD are denoted by blue triangles and local structures gathered from unstable material structures in OQMD are denoted by blue circles.

Figure \ref{fig:local_Nd} shows the correlation between Nd local structures and the structural stability in the t-SNE embedding space. In this map, we focus on three prominent groups denoted as $G_{Nd}^1$, $G_{Nd}^2$ and $G_{Nd}^3$. As the group with the highest number of local structures, $G_{Nd}^1$ holds local structures with fully occupied B neighbors. The number of B neighbor sites for local structures in this group is typically seven or eight. These B neighboring sites form several geometrical shapes surrounding the Nd center, as shown in the upper panel of  Fig. \ref{fig:local_Nd}. However, the most noticeable point is that there are no local structures in this group extracted from potentially formable structures (green or blue triangles). From this correlation, one may conclude that the potentially formable structures avoid Nd local structures containing fully occupied B neighboring sites as Nd local structures in this $G_{Nd}^1$ group.

In contrast, the groups $G_{Nd}^2$ and $G_{Nd}^3$  attract almost all Nd local structures gathered from potentially formable structures. By comparing the number of neighboring atoms,  Nd local structures in these  $G_{Nd}^2$ and $G_{Nd}^3$ groups generally have 12 to 20 neighboring atoms which is roughly twice times larger than Nd local structures in the $G_{Nd}^1$ group. In these groups,  the dominant geometrical shape of local structures is a pair of hexagonal pyramids, with the next Nd neighbor atoms located on top of each pyramid. The main difference among the local structures in this group is from the bases of these pyramids.  Since the pyramids’ bases of local structures in the group $G_{Nd}^2$ are only built from Fe atoms, the pyramids’ plates in the group $G_{Nd}^3$ are made entirely from B atoms. The middle of these hexagonal pyramids contains a hexagonal ring that comprises either Fe atoms, B atoms, or a combination of both atoms. From Fig. \ref{fig:local_Nd}, the most well-known magnet, Nd$_2$Fe$_{14}$B, has Nd local structures similar to the Nd local structures in the Nd$-2$ group. The main deviation is that the Nd local structure of Nd$_2$Fe$_{14}$B materials contain two hexagonal pyramids with Fe atoms locating on the two top. 

Figure \ref{fig:local_Fe} shows the correlation between Fe local structures and the structural stability in the t-SNE embedding space. In this Fe local structure distribution, there is a prominent group of Fe local structures denoted by $G_{Fe}^*$ in this figure. Similar to the Nd map, the elemental component of neighboring sites is the first criterion to identify local structures in the group $G_{Fe}^*$ and the other. The group $G_{Fe}^*$ contains local structures with fully extended B neighbor sites. The geometrical shapes of these B neighbors are hexahedrons, octahedrons, cuboctahedrons, etc. From right to left in the group Fe$^*$'s distribution, the number of B neighbor atoms surrounding the Fe local structures increases from 6 to 10.

In contrast,  local structures excluding  the group $G_{Fe}^*$  exhibit various arrangements of Nd, Fe, and B local atoms surrounding the central Fe sites. Their geometrical shapes are mostly built from rectangular pyramids with Nd atoms at the apex, and B or Fe neighbor atoms are in the base. As shown in Fig. \ref{fig:local_Fe}, from the bottom to the top of the region excluding the group $G_{Fe}^*$, one might notice that the  number of neighboring atoms in the local structures in this group increases from 9 to 12. In comparison, the number of neighboring atoms of the local structure in this region is significantly larger than local structures in the group $G_{Fe}^*$. 

The second difference between local structures in the $G_{Fe}^*$ group and the rest of the distribution is the stability of structures owing to these local structures. As shown in Fig. \ref{fig:local_Fe}, local structures in the $G_{Fe}^*$ group are entirely extracted from unstable structures (orange points). On the other hand, local structures  excluding the group $G_{Fe}^*$ contain a large number of local structure extracted from the potentially formable substituted structures (green points). This result is consistent with the correlation between the coordination number of Fe and the structure stability shown in Fig. \ref{fig:Distribution}  and Sect. 3.2. Figure \ref{fig:Distribution} shows that there is a higher possibility to obtain stable structures if the atomic coordination number of the Fe site ($d^6, s^2$) is larger than 6.5. Furthermore, all local structures collected from OQMD materials (blue points) do not also appear in the  $G_{Fe}^*$ group. Also from this Fig. \ref{fig:local_Fe}, the distribution Fe local structures collected from stable OQMD materials (blue triangles) appears in much consistent with the distribution of Fe local structures collected from potentially substituted structures (green triangles).  Finally, one might conclude that there is an undeniable correlation between the shape of Fe local structures and the stability of substituted Nd-Fe-B structures in our experiment. Potentially formable Nd-Fe-B substituted structures avoid possessing local Fe structures with entirely B atoms and low average coordination number.

Lastly, we investigated the distribution of B local structures as well as the correlation with structural stability through t-SNE embedding space in Fig. \ref{fig:local_B}. Since the distribution of the Fe and Nd local structures contains a small number of groups, i.e., two and four, respectively, the map of the B local structure comprises seven large groups. Intensively mining this map shows us the difference between the geometrical shapes of the local structures among these groups. From this figure, potentially formable structures appear within three main geometrical shapes of the B local structure shown in groups $G_{B}^1$, $G_{B}^2$ and $G_{B}^3$, respectively. The group $G_{B}^1$ possesses B local structures that are all in square-based pyramid form. These B local structures appear in Nd$_2$FeB$_{10}$, Nd$_4$FeB$_{14}$ and NdFeB$_{4}$-$\alpha$ substituted  structures, which possess B cage networks. B local structures in group $G_{B}^2$  appear in a B planar shape that extracts from NdFe$_2$B$_6$ or NdFeB$_{4}$-$\beta$ structures. Rather than forming B cage networks as B locals in $G_{B}^1$, B local structures in $G_{B}^1$ exhibit arrangements of mixing pentagonal and heptagonal planar rings. All noticeable forms of these B cage and planar network are visualized in Fig. 2 in the Supplementary section.

Lastly, local structures in $G_{B}^3$ appear with a limited number of neighboring B atoms. Fe atoms fully occupy almost all local structures in this group. In other words, structures possessing this type of local structure hold B atoms as interstitial sites rather than forming B networks as the others. Among B local structures extracted from  Nd-Fe-B structures in existing database OQMD, B local structures of well-known Nd$_2$Fe$_{14}$B magnets appear in this interstitial type. To conclude, in this study, we unveiled correlations between structural stability and Nd, Fe, and B local structures of both substituted Nd-Fe-B crystal structures and existed  Nd-Fe-B crystals from OQMD. These correlations provide a deep insight to  structure-stability relationship of Nd-Fe-B crystal structures that could be useful to speed up the screening process for the new formable crystal structures.

\section{4. Conclusion}

In this research, we examined the structure--stability relationship of substituted Nd-Fe-B crystal structures using descriptor-relevance analysis and the t-SNE dimensionality reduction method.  Hypothetical substituted Nd-Fe-B crystal structures are found using the elemental substitution method from LA-T-X host crystal structures with LA as lanthanide, T as transition metal, and X as light element (X = B, C, N, O). For each host crystal structure, a hypothetical crystal structure is created by substituting all lanthanide sites with Nd, all transition metal sites with Fe, and all light element sites with B. High-throughput first-principles calculations are used to evaluate the phase stability of the hypothetical crystal structures. Twenty of them are potentially formable. By performing the descriptor-relevance analysis on the orbital field matrix (OFM) materials' descriptor,  the average atomic coordination number is shown as the most essential essential factor that effects to the structure stability of these substituted Nd-Fe-B crystal structures. 19 among 20 hypothetical structures that are found potentially formable have an  average coordination number larger than 6.5. Under the use of the dimensionality reduction t-SNE method, concrete correlations between Nd, Fe and B local structures and the structural stability of these substituted crystals are revealed.   Unstable substituted structures frequently contain Nd and Fe local structures with a  low coordination number and fully occupied B neighboring atoms. On the other hand, potentially formable structures appear in three particular forms of B local structures: cage networks, planar networks, and interstitial. The extracted correlations between structural stability and different type of Nd, Fe and B local structures are validated with local structures of well known Nd-Fe-B crystals on OQMD. These structure-stability relationships are promising to accelerate the process of screening new formable Nd-Fe-B structures.

\section{Supplementary Materials}
See separated supplementary materials the results from descriptor relevance analysis and details information of 20 substituted structures that are potentially formable in our study.

\section*{acknowledgments}
\begin{acknowledgments}
This work is supported by the Ministry of Education, Culture, Sports, Science and Technology of Japan (MEXT) ESICMM Grant Number 12016013 and “Program for Promoting Researches on the Supercomputer Fugaku” (DPMSD), JSPS KAKENHI Grants Number 20K05301 and Number JP19H05815 (Grant-in-Aid for Scientific Research on Innovative Areas “Interface Ionics”). 
\end{acknowledgments}

\section*{Data availability}
The LATX dataset that support the findings of this study are openly available in  Open Quantum Materials Database (OQMD) \cite{OQMD} (version 1.1) at \href{http://oqmd.org/download/}{http://oqmd.org/download/}. Details of OQMD database are found in \href{https://doi.org/10.1007/s11837-013-0755-4}{https://doi.org/10.1007/s11837-013-0755-4}. Details of substituted NdFeB crystals that are potentially formable and the OQMD index of the corresponding host crystals could be found in Supplementary Materials. Three steps vasp calculations for Nd-Fe-B substituted structures are published at the NOMAD repository at \href{https://dx.doi.org/10.17172/NOMAD/2020.07.30-1}{\text{https://dx.doi.org/10.17172/NOMAD/2020.07.30-1}}. Summary of substituted structures and local structures, including vasp calculations, OFM descriptor are published in Zenodo at \href{https://doi.org/10.5281/zenodo.3966736}{https://doi.org/10.5281/zenodo.3966736}.

\bibliography{main}

\end{document}